\documentclass[aps,prb,floatfix,twocolumn]{revtex4}
\usepackage{graphicx}
\usepackage{amsmath}
\usepackage{amssymb}

\newcommand{\BEQ}{\begin{eqnarray}}
\newcommand{\EEQ}{\end{eqnarray}}
\newcommand{\BEA}{\begin{eqnarray}}
\newcommand{\EEA}{\end{eqnarray}}
\newcommand{\nn}{\nonumber}
\renewcommand{\d}{{\rm d}}

\begin{document} 
\draft
\title{Comment on ``Experimental violations of the second law of thermodynamics
for small systems and short timescales''}

\author{Theo M. Nieuwenhuizen$^{1}$ and Armen E. Allahverdyan$^{2}$}

\address{$^{1}$ Institute for Theoretical Physics, 
% University of Amsterdam\\
Valckenierstraat 65, 1018 XE Amsterdam, The Netherlands\\
$^{2}$
Yerevan Physics Institute,
Alikhanian Brothers St. 2, Yerevan 375036, Armenia}

\date{Version: July 22, 2002; today=\today}

%\maketitle

\begin{abstract}
The experimental verification of the Fluctuation Theorem by Wang et al.
is not a violation but even a confirmation of the second law, 
resulting from their observations in a proper interpretation.
\end{abstract}

\maketitle

In a recent paper Wang et al.~\cite{wang} consider a colloidal system and 
test the validity of the Fluctuation Theorem
\BEQ \frac{{\rm Pr}(\Sigma_t=A)}{{\rm Pr}(\Sigma_t=-A)}=\exp(A), \EEQ
where $\Sigma_t$ is the entropy production of a given trajectory of 
a colloidal particle, equal to the ratio of the dispersed energy and $k_BT$.

This equality is a strong statement about the second law, a field
often described by inequalities. It says that the probability for 
trajectories with positive entropy production is larger than the
one with negative production. As such, the theorem and the data of Wang
et al show that the {\it average} entropy production is positive.
Indeed, a small excercise yields from Eq. (1)
\BEQ \langle \Sigma_t\rangle&=&\int_{-\infty}^\infty\d A\, A~
{\rm Pr}(\Sigma_t=A)\\&=&
\int_0^\infty\d A\,A \,(1-e^{-A})\,{\rm Pr}(\Sigma_t=A) \nn
\EEQ
which is obviously positive. The positivity is also clear
from the date presented in ~\cite{wang}.

The statement made by the second law concerns the 
{\it average} entropy production, because the second law is related
to the {\it average} heat tranfer between system and bath.
Thus,  opposite to what the authors claim,
their data fully confirm the validity of the second law.

This insight in the meaning of the second law goes back to Maxwell,
who invented his so-called Maxwell demon with the motivation 
{\it to pick a hole in the second law of thermodynamics}, or, more 
precisely, {\it to show that the second law has only a statistical
nature}~\cite{klein,leffrex}. The aim of that excercise was of course
that by proving the non-existence of the demon, the second law in its
statistical nature would be proven.

It remains to stress that none of formulations of the second law known
to us ever claimed that unaveraged entropy production or unaveraged
work must be positive; see e.g. \cite{landau,balian,Len,Pusz,Boch,Kur,AN}.

We conclude that the statement ``violation of the second law'' made by
Wang et al is not an adequate term for the observed physics.

% This work is part of the research programme of the `Stichting 
% voor Fundamenteel Onderzoek der Materie'(FOM), which is financially
% supported by the `Nederlandse Organisatie voor Wetenschappelijk
% Onderzoek (NWO)'.


\begin{thebibliography}{99}

\bibitem{wang} G.M. Wang, E.M. Sevick, E. Mittag, D.J. Searles, and
D.J. Evans, Phys. Phys. Rev. Lett, {\bf 89}, 050601 (2002).

\bibitem{klein} Klein, M.J., {\it Maxwell, his demon, and the second law of 
thermodynamics}, Am. Sci. 58, 84-97 (1970); reprinted in ~\cite{leffrex}.

\bibitem{leffrex} Leff, H.S. and Rex, A.F. {\it Maxwell's demon: 
Entropy, Information, Computing}, (Adam Hilger, Bristol, 1990).

\bibitem{landau}L.D. Landau and E.M. Lifshitz, {\it Statistical
Physics, I}, (Pergamon Press Oxford, 1978).

\bibitem{balian}R. Balian, {\it From Microphysics to Macrophysics}, 
volumes I and II, (Springer, 1992).

\bibitem{Len} 
A. Lenard, J. Stat. Phys., {\bf 19}, 575, (1978).
\bibitem{Pusz} 
W. Pusz, L. Woronowicz, Comm. Math. Phys., {\bf 58}, 273, (1978). 
\bibitem{Boch} 
G.N. Bochkov and Yu.E. Kuzovlev, Physica A, {\bf 106}, (1981)
443; {\it ibid}, 480. 
\bibitem{Kur} 
J. Kurchan, cond-mat/0007360.

\bibitem{AN} 
A.E. Allahverdyan, and Th.M. Nieuwenhuizen,
Physica A {\bf 305}, (2002) 542.


\end{thebibliography}
\end{document}